\documentclass[reprint,
 amsmath,amssymb,
 aps,
]{revtex4-2}
\usepackage{graphicx}
\usepackage{dcolumn}
\usepackage{multirow}
\usepackage{bm}
\usepackage[table,xcdraw]{xcolor}
\usepackage{booktabs}
\usepackage{longtable}
\usepackage{blindtext}
\usepackage{ltablex}
\usepackage{appendix}
\usepackage[english]{babel}
\begin{document}
\title{Prediction of Alpha-Decay Half-Lives of Actinide Nuclei Using the DDM3Y Effective Interaction Potential}
\author{N.Sowmya$^{1,\ddagger}$, H.C.Manjunatha$^{2,\dagger}$, Roshini.K.N$^{1}$, R.S.Susheela$^{3,\& 4}$}
\affiliation{$^{1}$ Department of Physics, Govt. First Grade College, Chikkaballapur-562101, Karnataka, India}
\affiliation{ $^{2}$ Department of Physics, Govt. First Grade College, Devanahalli-562110, Karnataka, India}
\affiliation{$^{3}$Department of Physics, Government First Grade College, Nanjangud-571301, Mysore(D), India}
\affiliation{$^4$Department of Physics, R V College of Engineering, Bengaluru, Autonomous Institution affiliated to Visvesvaraya Technological University, Belagavi-590018, Karnataka, India}
\thanks {{}\\
Corresponding Author: 
$^\dagger$manjunathhc@rediffmail.com\\$\ddagger$sowmyaprakash8@gmail.com 
}
\begin{abstract}
\indent The prediction of nuclear half-lives is vital for understanding nuclear stability with significant applications in astrophysics, nuclear energy, and medical physics. This study investigates the $\alpha$-decay half-lives of 154 actinide nuclei in the atomic number range $89 \le Z \le 103$ using the Density-Dependent M3Y (DDM3Y) effective interaction potential. The theoretical framework utilizes a double-folding model where the densities of the $\alpha$-particle and the daughter nucleus are folded to derive the nuclear interaction potential.Theoretical half-lives were calculated using the WKB approximation and compared against experimental data and established semi-empirical models, including the Viola-Seaborg (VSS), CPPM, GLDM, and ELDM frameworks. The DDM3Y model demonstrates a systematically improved agreement with experimental half-lives across the actinide series, effectively capturing the inverse correlation between $Q$-values and decay times. Statistical analysis yielded a standard deviation of 1.76, confirming the reliability of this approach for predicting the stability and decay properties of heavy and new isotopes.
\end{abstract}
\maketitle
\section{Introduction}\label{introduction}
\indent Alpha decay ($\alpha$-decay) has been an essential tool in understanding nuclear forces and structure, offering critical insights into nuclear properties. Its importance extends to the identification and confirmation of newly synthesized elements, playing a key role in advancing nuclear science \cite{mang1964alpha}. This decay process continues to be invaluable, as evidenced by numerous studies \cite{hofmann2000discovery} that highlight its contribution to the exploration of nuclear phenomena and the discovery of superheavy elements. The theoretical modeling of alpha decay has evolved over decades, incorporating both microscopic and macroscopic approaches. \\
\indent {\cite{gurvitz1987decay} presented a two-potential approach for quasi-stationary state decay, yielding simple algebraic expressions for decay width and energy shift. In the quasiclassical limit, the width reduces to the Gamow formula with a defined pre-exponential factor. Further, $\alpha$-decay is a key process in heavy and superheavy nuclei, is analyzed using the two-potential approach with a quasistationary state approximation for even-even nuclei (Z = 62–118) \cite{sun2016systematic}. Incorporating isospin effects and hindrance factors improves agreement with experimental half-lives, matching results from cluster and liquid drop models.} Microscopic models, such as Relativistic Mean-Field (RMF) theory, have become powerful tools for describing nuclear ground-state properties, particularly those involving shell effects \cite{gambhir1990relativistic,reinhard1989relativistic,rutz1998odd}. Systematic predictions of alpha decay and spontaneous fission for superheavy nuclei in the range Z=106–118 have been carried out using the Hartree-Fock-Bogoliubov model with various Skyrme-type interactions \cite{seyyedi2020systematic}. 
Cluster radioactivity, an intermediate process between alpha decay and spontaneous fission, was studied in proton-rich Osmium isotopes (A=162–190) using the unified fission model and Hartree–Fock–Bogoliubov theory \cite{ashok2016cluster}. \citet{lovas1998microscopic} studied microscopic theory of cluster radioactivity. The structure of the newly discovered nuclides, including $^{259}$Db, from the $^{241}$Am($^{22}$Ne,4n) reaction and its $\alpha$-decay chain, is analyzed using the relativistic mean field (RMF) approach with the NL3 and TM1 effective interactions \cite{long2002structure}. On the other hand, macroscopic models, such as the generalized liquid drop model \cite{dasgupta2007generalized,hong2009alpha}, Coulomb and proximity potential \cite{yao2015comparative,myers2000nucleus,sowmya2020competition,sowmya2023radioactive,manjunatha2017study}, effective liquid drop model \cite{cui2022improved,sridhara2021study,goncalves1993effective,abdulla2024systematic,cui2020two} focus on alpha-decay, cluster and proton radioactivity. Semi-empirical formulas, such as the Viola-Seaborg formula \cite{viola1966nuclear}, Royer formula \cite{royer2000alpha},  Brown formula \cite{brown1992simple}, Denisov and Khudenko \cite{denisov2009alpha}, Horoi et al., \cite{horoi2004scaling}, Ni et al.,\cite{ni2008unified}, Qi et al., \cite{qi2009universal} provide convenient and accurate predictions of alpha-decay half-lives by correlating them with decay energy (Q$_{\alpha}$ ) and atomic numbers. These approaches bridge theoretical models with experimental data, enabling robust predictions for nuclei far from stability.\\
\indent The density dependence of effective nucleon-nucleon (NN) interactions was investigated through double-folding model analyses of elastic alpha-particle scattering \cite{gils1987density}. Further, cluster-decay half-lives of heavy nuclei \cite{yahya2022cluster}, $\alpha$-decay half-lives in the atomic number range $52\le Z\le 110$ \cite{xu2006global},
Nucleus–Nucleus potentials were determined using the double-folding model with M3Y–Reid and M3Y–Paris effective nucleon–nucleon (NN) interactions \cite{gao2009nucleon}, considering both zero-range and finite-range exchange parts. For the spherical projectile-target system $^{16}$O+$^{208}$Pb, the fusion cross sections, barrier energies, and barrier distributions were evaluated. The radioactive decay of nuclei through $\alpha$ particle emission has been theoretically studied within the framework of a super asymmetric fission model, utilizing the double folding (DF) method to calculate the $\alpha$-nucleus interaction potential.\\
\indent Motivated by these advancements, this study seeks to investigate the alpha-decay half-lives of actinide nuclei in the atomic number range $89 \le Z\le 103$ using nuclear potentials derived from the M3Y effective interaction (DDM3Y). By folding the densities of the alpha particle and the daughter nucleus, this study aims to improve the accuracy of alpha-decay predictions in this atomic region, offering new insights into the stability and decay properties of actinide nuclei. The results obtained using DDM3Y will be compared with existing semi-empirical formulas to assess the effectiveness of the model and its potential for predicting new isotopes in the actinide series. 
\section{Theoretical Framework}
\begin{figure}[!ht]
    \centering
\includegraphics[width=\linewidth]{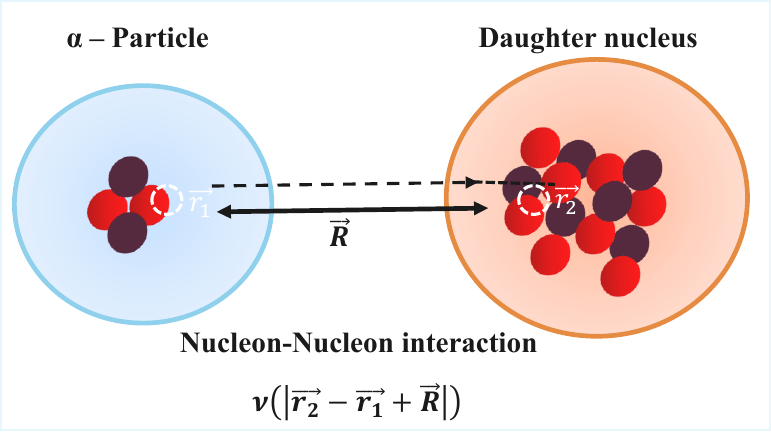}
\caption{Schematic of the DDM3Y double-folding model for the $\alpha$-daughter system. Here, $\vec{r}_1$ and $\vec{r}_2$ denote the positions of nucleons in the $\alpha$ particle and daughter nucleus, respectively, and $\vec{R}$ is the center-to-center separation. The effective interaction $\nu(|\vec{r}_2-\vec{r}_1+\vec{R}|)$ is folded over the two densities to obtain the $\alpha$-daughter potential.}\label{density}
\end{figure}

\indent The effective $\alpha$-nucleus potential $V(R)$ in the atomic number range 89$\le Z\le $103 is given by \cite{chowdhury2006alpha};
\begin{align}
    V(R)= V_N(R)+V_C(R)+\frac{\hbar^2 \ell(\ell+1)}{2\mu R^2}
\end{align}
here $V_C(R)$ is the Coulomb potential, $V_N$ is the double folded nuclear interaction potential, and $\frac{\hbar^2 \ell(\ell+1)}{2\mu R^2}$ is the centrifugal potential. Here $\mu$ is the reduced mass of the masses of the emitted particle and  the daughter nuclei. $\ell$ is the angular momentum and in the present work, we have considered ground-state to ground-state transitions of the nuclei. The Coulomb potential is expressed as;
\begin{align}
    V_C(R)=Z_1Z_2e^2\begin{cases}
        \frac{1}{R} \;\;\;\;\;\;\;\;\;\;\;\;\;\;\;\;\;\;\;\;\;\;\;\;\;\; \text{for}\;\;  R>R_C\\
        \frac{1}{2R_C}\left[3-\left(\frac{R}{R_C}\right)^2\right]  \text{for } R\le R_C
    \end{cases}
\end{align}
where $Z_1$ and $Z_2$ are the atomic number of $\alpha$ and daughter nucleus respectively and $R_C=1.2 (A_1^{1/3}+A_2^{1/3})$fm.\\
The double-folded nuclear interaction potential between the emitted particle and the daughter nucleus is described as follows \cite{satchler1979folding}:
\begin{align}
    V_N(R)=\int \int \rho_1(\vec{r_1}) \rho_2(\vec r_2) \upsilon [|\vec{r_2}-\vec{r_1}+\vec{R}|]d^3r_1 d^3r_2 \label{eq1}
\end{align}
Here, $\rho_1$ and $\rho_2$ represent the density distribution functions of the two nuclear fragments. Further, $\vec{r}_1$ and $\vec{r}_2$ denote the positions of nucleons in the $\alpha$-particle and daughter nucleus, respectively, and $\vec{R}$ is the center-to-center separation as shown in figure \ref{density}. 
For the $\alpha$-particle, the density distribution function is expressed in a Gaussian form;
\begin{align}
    \rho(r)=0.4229\,\exp(-0.7024 r^2)\label{equation-4}
\end{align}
the volume integral of the $\alpha$-particle density equals $A_\alpha$=4, its mass number. For heavier nuclei, experimental charge density distributions are well-represented by a two-parameter Fermi function. Since proton and neutron densities share similar forms due to comparable nuclear forces, the daughter nucleus's matter density follows a spherically symmetric Fermi function;
\begin{align}
    \rho(r)=\rho_0/[1+exp((r-c)/a)]\label{rho}
\end{align}
where $r_i=1.13 A_i^{1/3}$ is the sharp radius, $c=r_i(1-\pi^2a^2/3r_i^2)$ is the half-density, $a=0.54$fm \cite{srivastava1974effects} is the diffuseness parameter. $\rho_0$ is the central density is determined by ensuring that the volume integral of the density distribution function equals the mass number $A_d$ of the residual daughter nucleus. In equation \ref{eq1}, $\upsilon [|\vec{r_2}-\vec{r_1}+\vec{R}|]$ where $s=|\vec r_2-\vec r_1+\vec R|$, with $\upsilon (s, \rho_1, \rho_2, E)$ is expressed as;
\begin{align}
    \upsilon (s, \rho, E)=t^{M3Y}(s, E)g(\rho_1, \rho_2, E)\label{eq2}
\end{align}
where the density dependence term $g(\rho_1, \rho_2, E)$ has now been factorized into a target term times a projectile term \cite{srivastava1983density};
\begin{align}
    g(\rho_1, \rho_2, E)=C(1-\beta(E)\rho_1^{2/3})(1-\beta(E)\rho_2^{2/3}).
\end{align}
The folding model potentials, derived by double-folding the density distributions $\rho_1$ of the $\alpha$-particle and $\rho_2$ of the daughter nucleus using the factorized density-dependent M3Y-Reid-Elliott effective interaction, combined with a zero-range potential for single-nucleon exchange, have been effectively utilized to estimate $\alpha$-decay half-lives for newly synthesized elements and their isotopes. The term $t^{M3Y}$ is the M3Y effective interaction potential is expressed as;
\begin{align}
    t^{M3Y}=9968\frac{e^{-3.97r}}{4r}-6661\frac{e^{-2.79r}}{4r}+J_{00}(E)\delta(r)\label{eq_T_M3Y}
\end{align}
here $J_{00}$ is the zero-range pseudopotential for
single-nucleon exchange and is given by;
\begin{align}
    J_{00}(E)=-276(1-0.005E/A_\alpha) \text(MeV. fm^3)
\end{align}
$J_{00} (E)$ remains nearly constant for $\alpha$-decay processes and can be approximated as -276 MeV$fm^3$ \cite{chowdhury2006alpha}. However, in the present we have considered it as zero. The amount of energy released during $\alpha$-decay is expressed as;
\begin{align}
    Q=[M-M_\alpha-M_d]c^2\label{q-value}
\end{align}
The $\alpha$-decay is energetically feasible when the Q-value is positive. Here $M$, $M_\alpha$, and $M_d$ are the mass excess values which have been taken from the available mass excess data \cite{moller2016nuclear,audi2003ame2003,RIPL-3}. The $\alpha$-decay half-life is given by;
\begin{align}
    T_{1/2}=\frac{ln 2}{\lambda}=\frac{ln 2}{\nu P P_\alpha}\label{eq11}
\end{align}
here, $P_\alpha$ is the pre-formation probability, where $P_{\alpha}$=0.43 for even–even nuclei, $P_\alpha$=0.35 for odd–A nuclei, and $P_\alpha$=0.18 for
odd–odd nuclei \cite{xu2005favored}. The number of assaults on the barrier per second $\nu=\frac{\omega}{2\pi}=\frac{2E_\nu}{h}$, and $\lambda$ is the decay constant and it is expressed as;
\begin{align}
    E_\nu=Q\left(0.056+0.039 exp\left[\frac{4-A_2}{2.5}\right]\right)
\end{align}
in equation \ref{eq11}, $P$ \cite{ahmed2013clusterization,deng2016alpha,deng2018systematic} is the barrier penetrability and it is expressed as 
\begin{align}
    P=exp\left(-\frac{2}{\hbar}\int_a^b \sqrt{2\mu(V(R)-Q)}dr\right)\label{penetration}
\end{align}
here $a$ and $b$ are the turning points of the WKB integral determined from equation $V(a)=V(b)=Q$.
\section*{VIOLA-SEABORG SEMI-EMPIRICAL FORMULA}
\indent Viola and Seaborg proposed a simple formula based on the Gamow model, which yields the logarithmic $\alpha$-decay half-lives by incorporating intercept parameters that exhibit a linear dependence on the charge number of the daughter nucleus. The corresponding expression is given below\cite{viola1966nuclear}.
\begin{equation}
\log_{10} T_{1/2}({s}) =
\frac{a Z + b}{\sqrt{Q}} + c Z + d+h,
\end{equation}
The present paper considered fitted parameters are, $a=1.3892$, $b=-13.862$, $c=-0.1086$, $d=-41.458$
$h_{ee}=0$, $h_{eo}=0.641$, $h_{oe}=0.437$ and $h_{oo}=1.024$.
\section{Results and Discussions}\label{results}
\begin{figure}[!ht]
    \centering
\includegraphics[width=\linewidth]{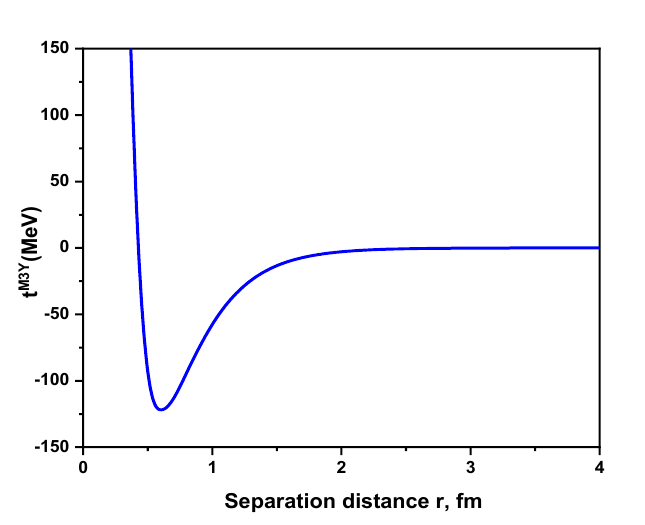}
\caption{A plot of M3Y effective interaction potential versus separation distance.}\label{figure-3}
\end{figure}
\begin{figure}[!ht]
    \centering
\includegraphics[width=\linewidth]{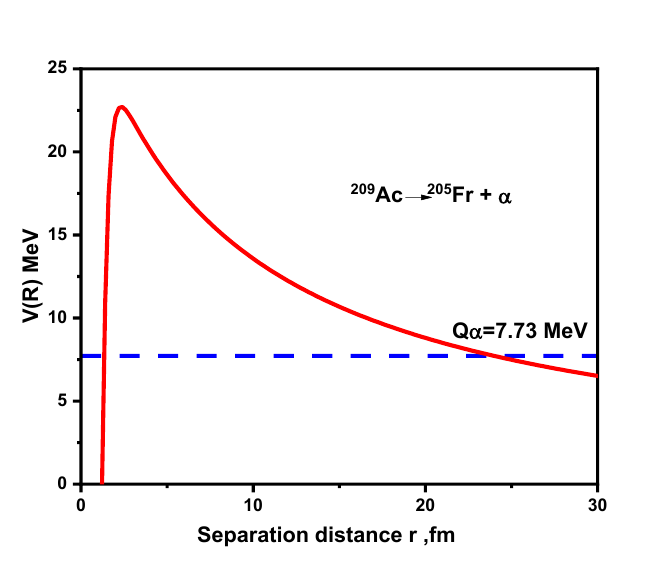}
\caption{A plot of total potential as a function of separation distance for $\alpha$-decay from the parent nuclei $^{209}$Ac nuclei.}\label{figure-2}
\end{figure}

\indent  The amount of energy released during $\alpha$-decay in the atomic number range $89\le Z\le 103$ have been evaluated using equation \ref{q-value}. We have considered all isotopes of each atomic number with positive Q-value. Further, the density distributions of $\alpha$ and daughter nuclei is obtained using equations \ref{equation-4} and \ref{rho}. Further, the M3Y effective interaction potential $t^{M3Y}$ is evaluated using equation \ref{eq_T_M3Y}. The $t^{M3Y}$ effective interaction potential exhibits a characteristic behavior as a function of separation distance as seen in figure \ref{figure-2}. The figure illustrates the variation of the $t^{M3Y}$ interaction potential  as a function of separation distance $r$(in fm). The potential is strongly attractive at short distances, reaching large negative values, which reflects the dominance of nuclear forces when interacting particles are close. As the separation distance increases, the potential rapidly rises toward zero, indicating a weakening interaction as nuclear forces become negligible. The smooth curve suggests a well-behaved effective interaction model, consistent with realistic nucleon–nucleon potentials. This behavior highlights the short-range nature of nuclear forces and the transition from strong attraction at small $r$ to near-zero interaction at larger distances.\\
\indent Further, the total potential is evaluated using equation (1). All calculations were carried out assuming zero angular momentum transfer i.e $\ell=0$. Hence the centrifugal potential term $\approx$0. The double-folded nuclear interaction potential is evaluated \ref{eq1}, and the Coulomb potential is calculated using equation (2). Figure \ref{figure-2} shows a plot of total potential which combines nuclear, Coulomb, and centrifugal (with $\ell=0$) components as a function of separation distance for $\alpha$-decay from the parent nuclei $^{209}$Ac nuclei. The barrier height and width significantly influence the decay rate. Higher barriers correspond to lower tunneling probabilities and longer half-lives. Furthermore, the barrier penetrability is evaluated using WKB approximation within the boundary conditions as detailed in equation (\ref{penetration}) and half-life using equation (\ref{eq11}). \\
\indent The evaluated half-lives from the DDM3Y model has been compared with the available experimental data in the atomic number range $89\le Z\le 103$ \cite{ZHU20211,SINGH20191,SINGH2019405,BROWNE20111115,KONDEV20081527,kondev2011707,MARTIN20071583,CHEN2015373,SINGH2013661,AURANEN2020117,SINGH20132023,WU20071057,KONDEV2018382,KUMARJAIN2007883,Jain:2009yef,KONDEV2016257,ABUSALEEM2014163,BROWNE20082657,BROWNE20062579,SINGH2020499,Browne:2006zz,BROWNE2014205,BROWNE2015191,BROWNE2014293,SINGH20082439,Nesaraja:2017ycm,NESARAJA2015183,cdb3cef5313c453cbc6af44608ec0b60,MARTIN2014377,BROWNE20131041,SINGH2017327,Singh:2017wcl,Mattera:2021mgo,NESARAJA2015395,BROWNE20111833,SHAMSUZZOHA2014nuclear,M.S.Basunia2006nuclear,AKOVALI2001131}.  This comparison is tabulated in table \ref{table_comparison} and its continuation table. In addition, table shows a comparison of $\alpha$-decay half-lives obtained from the present work (PW) with experimental measurements and predictions from established theoretical and semi-empirical models, including CPPM \cite{seyyedi2020systematic}, GLDM \cite{royer2000alpha}, ELDM \cite{oap1998effective,sridhara2021study} and semi empirical model such as VSS \cite{viola1966nuclear}. Across a broad range of parent nuclei spanning Ac to Lr isotopes, the present model demonstrates a systematically improved agreement with experimental data. The PW results successfully reproduce the expected inverse correlation between Q-value and half-life, confirming that the model effectively captures the barrier penetration dynamics influencing $\alpha$-emission. In contrast, the VSS formula frequently underestimates half-lives, likely due to its reliance on global parameterizations. Similarly, CPPM tends to overestimate decay times, reflecting limitations in its treatment of proximity interactions and Coulomb barrier shaping. However, GLDM and ELDM yield competitive predictions for selected heavy nuclei. However, the PW maintains stable accuracy in both even–even and odd-mass nuclei.\\
\indent While Table \ref{table_comparison} provides a comprehensive comparison between experimental $\alpha$-decay half-lives and the results of the present work alongside other established models, a quantitative statistical analysis is required to draw definitive conclusions. To this end, we evaluated the root-mean-square deviation ($\sigma$) for the $N=154$ actinide isotopes using the following expression:
\begin{align}
\sigma = \sqrt{\frac{1}{N-1} \sum_{i=1}^{N} \left( \log_{10} T_{1/2, i}^{\text{EXP}} - \log_{10} T_{1/2, i}^{\text{PW}} \right)^2}
\end{align}
where $\log_{10} T_{1/2}^{\text{EXP}}$ and $\log_{10} T_{1/2}^{\text{PW}}$ represent the experimental and theoretical logarithmic half-lives, respectively. The calculated standard deviation for the present model is 1.76, indicating a strong correlation with the experimental data across the actinide region.\\
\begin{figure}[!ht]
    \centering
\includegraphics[width=\linewidth]{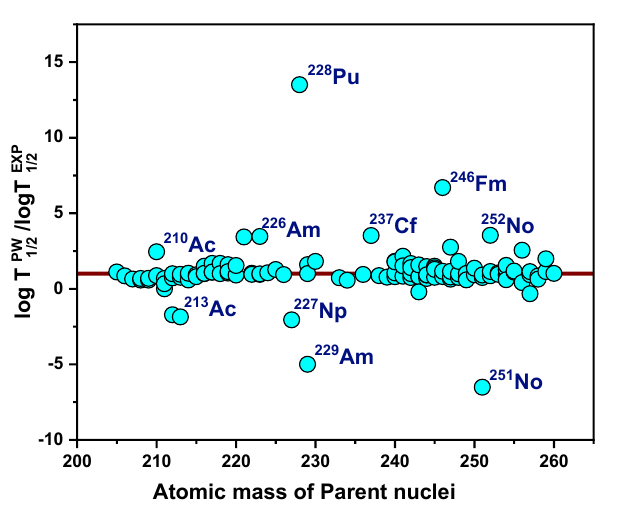}
\caption{A plot of log$T_{1/2}^{PW}$/log$T_{1/2}^{EXP}$ as a function of atomic mass of parent nuclei in the range 200 to 260. }\label{figure-4}
\end{figure}
\indent Further, the figure \ref{figure-4} provided illustrates the deviations between the experimental half-lives of $\alpha$-decay and the theoretical predictions of the present work (PW) for actinide nuclei across an atomic mass range of approximately 200 to 260. The vertical axis represents the ratio of $\log T_{1/2}^{\text{PW}}$  to the $\log T_{1/2}^{\text{EXP}}$, which serves as a metric for the model's accuracy. A significant majority of the data points are densely clustered around the horizontal line $\approx 1$, indicating that the DDM3Y effective interaction potential provides a high level of agreement with experimental values for most isotopes in this region. However, the figure also identifies several notable outliers where theoretical predictions deviate more substantially from experimental observations. Notable isotopes with higher positive deviations include $^{228}\text{Pu}$, $^{246}\text{Fm}$, and $^{252}\text{No}$, while significant negative deviations are visible for $^{213}\text{Ac}$, $^{227}\text{Np}$, $^{229}\text{Am}$, and $^{251}\text{No}$. These discrepancies may arise from specific nuclear structures, such as shell effects or deformation, which are particularly pronounced in certain actinide isotopes. Overall, the plot effectively visualizes the model's systematic performance and highlights specific areas for potential theoretical refinement.
\begin{table*}[htbp]
\caption{A comparison of half-lives obtained using present work with those of experiments, and theoretical models such as Coulomb and Proximity Potential Model (CPPM), Generalized liquid drop model (GLDM), effective liquid drop model (ELDM) and semi-empirical formula such as Viola and Seaborg (VSS).} \label{table_comparison}
\centering
\begin{tabular}{|c|c|cccccc|}
\hline
\multirow{2}{*}{\textbf{\begin{tabular}[c]{@{}c@{}}Parent \\ Nuclei\end{tabular}}} & \multirow{2}{*}{\textbf{Q(MeV)\cite{ENSDF}}} & \multicolumn{6}{c|}{\textbf{Half-lives}} \\ \cline{3-8} 
 &  & \multicolumn{1}{c|}{\textbf{EXP}} & \multicolumn{1}{c|}{\textbf{PW}} & \multicolumn{1}{c|}{\textbf{VSS\cite{viola1966nuclear}}} & \multicolumn{1}{c|}{\textbf{CPPM\cite{seyyedi2020systematic}}} & \multicolumn{1}{c|}{\textbf{GLDM\cite{royer2000alpha}}} & \textbf{ELDM\cite{oap1998effective,sridhara2021study}
 } \\ \hline
$^{205}$Ac & 8.093$\pm$0.059 & \multicolumn{1}{c|}{$20^{+97}_{-9}ms$} & \multicolumn{1}{c|}{$12.72{ms}$} & \multicolumn{1}{c|}{$1.62{ms}$} & \multicolumn{1}{c|}{$38.7{ms}$} & \multicolumn{1}{c|}{-} & - \\
$^{206}$Ac & 7.958$\pm$0.065 & \multicolumn{1}{c|}{$22^ {+9}_{-5}ms$} & \multicolumn{1}{c|}{$37.70{ms}$} & \multicolumn{1}{c|}{$4.16{ms}$} & \multicolumn{1}{c|}{$103{ms}$} & \multicolumn{1}{c|}{-} & - \\
$^{207}$Ac & 7.845$\pm$0.056 & \multicolumn{1}{c|}{$27^{+11}_{-6}ms$} & \multicolumn{1}{c|}{$96.17{ms}$} & \multicolumn{1}{c|}{$9.30{ms}$} & \multicolumn{1}{c|}{$242{ms}$} & \multicolumn{1}{c|}{-} & - \\
$^{208}$Ac & 7.729$\pm$0.060 & \multicolumn{1}{c|}{$95^{+24}_{-16}ms$} & \multicolumn{1}{c|}{$259.24{ms}$} & \multicolumn{1}{c|}{$21.6{ms}$} & \multicolumn{1}{c|}{$587{ms}$} & \multicolumn{1}{c|}{-} & - \\
$^{209}$Ac & 7.73$\pm$0.055 & \multicolumn{1}{c|}{$0.087^{+12}_{-9}s$} & \multicolumn{1}{c|}{$245.87{ms}$} & \multicolumn{1}{c|}{$0.022{s}$} & \multicolumn{1}{c|}{$0.578{s}$} & \multicolumn{1}{c|}{$0.033{s}$} & $0.075{s}$ \\
$^{210}$Ac & 7.586$\pm$0.057 & \multicolumn{1}{c|}{$0.35 ^{+5}_{-5}s$} & \multicolumn{1}{c|}{$76.58{ms}$} & \multicolumn{1}{c|}{$0.063{s}$} & \multicolumn{1}{c|}{$0.002{s}$} & \multicolumn{1}{c|}{$0.087{s}$} & $0.180{s}$ \\
$^{211}$Ac & 7.568$\pm$0.052 & \multicolumn{1}{c|}{$0.21^{+3}_{-3}s$} & \multicolumn{1}{c|}{$0.995{s}$} & \multicolumn{1}{c|}{$0.072{s}$} & \multicolumn{1}{c|}{$0.003{s}$} & \multicolumn{1}{c|}{$0.072{s}$} & $0.150{s}$ \\
$^{212}$Ac & 7.54$\pm$0.024 & \multicolumn{1}{c|}{$880^{+35}_{35}ms$} & \multicolumn{1}{c|}{$1.246{s}$} & \multicolumn{1}{c|}{$89.4{ms}$} & \multicolumn{1}{c|}{$370{ms}$} & \multicolumn{1}{c|}{$155{ms}$} & $300{ms}$ \\
$^{213}$Ac & 7.498$\pm$0.004 & \multicolumn{1}{c|}{$738^{+16}_{-16}ms$} & \multicolumn{1}{c|}{$1.758{s}$} & \multicolumn{1}{c|}{$123{ms}$} & \multicolumn{1}{c|}{$5{ms}$} & \multicolumn{1}{c|}{$174{ms}$} & $360{ms}$ \\
$^{214}$Ac & 7.351$\pm$0.025 & \multicolumn{1}{c|}{$8.2^{+2}_{-2}s$} & \multicolumn{1}{c|}{$6.918{s}$} & \multicolumn{1}{c|}{$0.390{s}$} & \multicolumn{1}{c|}{$0.18{s}$} & \multicolumn{1}{c|}{$0.562{s}$} & $0.525{s}$ \\
$^{208}$Th & 8.202$\pm$0.031 & \multicolumn{1}{c|}{$1.7^{+17}_{-6}ms$} & \multicolumn{1}{c|}{$12.45{ms}$} & \multicolumn{1}{c|}{$1.84{ms}$} & \multicolumn{1}{c|}{$40.9{ms}$} & \multicolumn{1}{c|}{-} & - \\
$^{209}$Th & 8.166$\pm$0.106 & \multicolumn{1}{c|}{$2.5^{+17}_{-7}ms$} & \multicolumn{1}{c|}{$16.03{ms}$} & \multicolumn{1}{c|}{$2.35{ms}$} & \multicolumn{1}{c|}{$51.3{ms}$} & \multicolumn{1}{c|}{-} & $1.41{ms}$ \\
$^{210}$Th & 8.069$\pm$0.006 & \multicolumn{1}{c|}{$16^{+4}_{-4}ms$} & \multicolumn{1}{c|}{$27.70{ms}$} & \multicolumn{1}{c|}{$4.60{ms}$} & \multicolumn{1}{c|}{$106{ms}$} & \multicolumn{1}{c|}{-} & $17{ms}$ \\
$^{211}$Th & 7.937$\pm$0.063 & \multicolumn{1}{c|}{$37 ^{+28}_{-11}ms$} & \multicolumn{1}{c|}{$104.17{ms}$} & \multicolumn{1}{c|}{$11.7{ms}$} & \multicolumn{1}{c|}{$279{ms}$} & \multicolumn{1}{c|}{-} & $13.8{ms}$ \\
$^{212}$Th & 7.958$\pm$0.005 & \multicolumn{1}{c|}{$31.7^{+13}_{-13}ms$} & \multicolumn{1}{c|}{$82.90{ms}$} & \multicolumn{1}{c|}{$10.1{ms}$} & \multicolumn{1}{c|}{$238{ms}$} & \multicolumn{1}{c|}{-} & $31{ms}$ \\
$^{213}$Th & 7.837$\pm$0.007 & \multicolumn{1}{c|}{$146^{+22}_{-19}ms$} & \multicolumn{1}{c|}{$230.86{ms}$} & \multicolumn{1}{c|}{$24.0{ms}$} & \multicolumn{1}{c|}{$763{ms}$} & \multicolumn{1}{c|}{$32.4{ms}$} & $71{ms}$ \\
$^{214}$Th & 7.827$\pm$0.005 & \multicolumn{1}{c|}{$87^{+10}_{-10}ms$} & \multicolumn{1}{c|}{$241.11{ms}$} & \multicolumn{1}{c|}{$25.9{ms}$} & \multicolumn{1}{c|}{$877{ms}$} & \multicolumn{1}{c|}{$33.1{ms}$} & $73{ms}$ \\
$^{216}$Th & 8.072$\pm$0.004 & \multicolumn{1}{c|}{$26.0^{+2}_{-2}ms$} & \multicolumn{1}{c|}{$26.00{ms}$} & \multicolumn{1}{c|}{$4.50{ms}$} & \multicolumn{1}{c|}{$144{ms}$} & \multicolumn{1}{c|}{$5.25{ms}$} & $11{ms}$ \\
$^{217}$Th & 9.435$\pm$0.004 & \multicolumn{1}{c|}{$0.252^{+4}_{-4}ms$} & \multicolumn{1}{c|}{$0.906{\mu s}$} & \multicolumn{1}{c|}{$0.966{ms}$} & \multicolumn{1}{c|}{$0.013.9{ms}$} & \multicolumn{1}{c|}{-} & $0.692{ms}$ \\
$^{211}$Pa & 8.481$\pm$0.041 & \multicolumn{1}{c|}{$> 300 ns$} & \multicolumn{1}{c|}{$313.1{ms}$} & \multicolumn{1}{c|}{$374{ns}$} & \multicolumn{1}{c|}{$13.1{ms}$} & \multicolumn{1}{c|}{-} & - \\
$^{212}$Pa & 8.411$\pm$0.059 & \multicolumn{1}{c|}{$5.1^{+51}_{-17}ms$} & \multicolumn{1}{c|}{$5.231{ms}$} & \multicolumn{1}{c|}{$1.07{ms}$} & \multicolumn{1}{c|}{$20.6{ms}$} & \multicolumn{1}{c|}{-} & - \\
$^{213}$Pa & 8.384$\pm$0.012 & \multicolumn{1}{c|}{$5.3^{+40}_{-16}ms$} & \multicolumn{1}{c|}{$6.235{ms}$} & \multicolumn{1}{c|}{$1.28{ms}$} & \multicolumn{1}{c|}{$24.2{ms}$} & \multicolumn{1}{c|}{$1.51{ms}$} & $340{ms}$ \\
$^{214}$Pa & 8.271$\pm$0.052 & \multicolumn{1}{c|}{$5.3^{+40}_{-16}ms$} & \multicolumn{1}{c|}{$15.03{ms}$} & \multicolumn{1}{c|}{$2.73{ms}$} & \multicolumn{1}{c|}{$54.3{ms}$} & \multicolumn{1}{c|}{-} & $7.4{ms}$ \\
$^{215}$Pa & 8.236$\pm$0.063 & \multicolumn{1}{c|}{$14^{+2}_{-2}ms$} & \multicolumn{1}{c|}{$19.19{ms}$} & \multicolumn{1}{c|}{$3.47{ms}$} & \multicolumn{1}{c|}{$77.4{ms}$} & \multicolumn{1}{c|}{$6.46{ms}$} & $8.7{ms}$ \\
$^{216}$Pa & 8.099$\pm$0.011 & \multicolumn{1}{c|}{$0.15^{+6}_{-4}s$} & \multicolumn{1}{c|}{$58.75{ms}$} & \multicolumn{1}{c|}{$0.009{s}$} & \multicolumn{1}{c|}{$0.247{s}$} & \multicolumn{1}{c|}{-} & $0.024{s}$ \\
$^{217}$Pa & 8.489$\pm$0.004 & \multicolumn{1}{c|}{$3.8^{+2}_{-2}ms$} & \multicolumn{1}{c|}{$2.253{ms}$} & \multicolumn{1}{c|}{$0.640{ms}$} & \multicolumn{1}{c|}{$14.6{ms}$} & \multicolumn{1}{c|}{$0.617{ms}$} & $1.4{ms}$ \\
$^{218}$Pa & 9.791$\pm$0.012 & \multicolumn{1}{c|}{$109^{+5}_{-5}\mu s$} & \multicolumn{1}{c|}{$0.216{\mu s}$} & \multicolumn{1}{c|}{$0.309{\mu s}$} & \multicolumn{1}{c|}{$3.78{\mu s}$} & \multicolumn{1}{c|}{-} & $0.63{\mu s}$ \\
$^{219}$Pa & 10.128$\pm$0.069 & \multicolumn{1}{c|}{$54^{+10}_{-10}ns$} & \multicolumn{1}{c|}{$26.19{ns}$} & \multicolumn{1}{c|}{$54{ns}$} & \multicolumn{1}{c|}{$0.636{\mu s}$} & \multicolumn{1}{c|}{-} & $0.140{\mu s}$ \\
$^{220}$Pa & 9.704$\pm$0.011 & \multicolumn{1}{c|}{$0.78^{+16}_{-16}\mu s$} & \multicolumn{1}{c|}{$0.345{\mu s}$} & \multicolumn{1}{c|}{$0.491{\mu s}$} & \multicolumn{1}{c|}{$6.78{\mu s}$} & \multicolumn{1}{c|}{-} & $0.480{\mu s}$ \\
$^{221}$Pa & 9.248$\pm$0.058 & \multicolumn{1}{c|}{$5.9^{+17}_{-17}\mu s$} & \multicolumn{1}{c|}{$6.829{\mu s}$} & \multicolumn{1}{c|}{$6.14{\mu s}$} & \multicolumn{1}{c|}{$122{\mu s}$} & \multicolumn{1}{c|}{-} & $11{\mu s}$ \\
$^{215}$U & 8.588$\pm$0.059 & \multicolumn{1}{c|}{$0.7^{+13}_{-3}ms$} & \multicolumn{1}{c|}{$3.058{ms}$} & \multicolumn{1}{c|}{$0.782{ms}$} & \multicolumn{1}{c|}{$15.1{ms}$} & \multicolumn{1}{c|}{-} & - \\
$^{216}$U & 8.531$\pm$0.026 & \multicolumn{1}{c|}{$4.5^{+47}_{-16}ms$} & \multicolumn{1}{c|}{$4.605{ms}$} & \multicolumn{1}{c|}{$1.13{ms}$} & \multicolumn{1}{c|}{$22.1{ms}$} & \multicolumn{1}{c|}{-} & - \\
$^{217}$U & 8.426$\pm$0.080 & \multicolumn{1}{c|}{$16^{+21}_{-6}ms$} & \multicolumn{1}{c|}{$10.26{ms}$} & \multicolumn{1}{c|}{$2.27{ms}$} & \multicolumn{1}{c|}{$50.9{ms}$} & \multicolumn{1}{c|}{-} & - \\
$^{218}$U & 8.775$\pm$0.009 & \multicolumn{1}{c|}{$0.65^{+8}_{-7}ms$} & \multicolumn{1}{c|}{$0.626{ms}$} & \multicolumn{1}{c|}{$0.237{ms}$} & \multicolumn{1}{c|}{$0.408{ms}$} & \multicolumn{1}{c|}{-} & $510{\mu s}$ \\
$^{219}$U & 9.95$\pm$0.012 & \multicolumn{1}{c|}{$60^{+7}_{-7}\mu s$} & \multicolumn{1}{c|}{$0.186{\mu s}$} & \multicolumn{1}{c|}{$0.290{\mu s}$} & \multicolumn{1}{c|}{$3.59{\mu s}$} & \multicolumn{1}{c|}{-} & - \\
$^{222}$U & 9.481$\pm$0.051 & \multicolumn{1}{c|}{$4.6^{+7}_{-7}\mu s$} & \multicolumn{1}{c|}{$3.382{\mu s}$} & \multicolumn{1}{c|}{$3.63{\mu s}$} & \multicolumn{1}{c|}{$59.2{\mu s}$} & \multicolumn{1}{c|}{-} & $6.1{\mu s}$ \\
$^{223}$U & 9.158$\pm$0.017 & \multicolumn{1}{c|}{$18^{+10}_{-5}\mu s$} & \multicolumn{1}{c|}{$30.15{\mu s}$} & \multicolumn{1}{c|}{$23.1{\mu s}$} & \multicolumn{1}{c|}{$488{\mu s}$} & \multicolumn{1}{c|}{-} & $160{\mu s}$ \\
$^{219}$Np & 9.207$\pm$0.041 & \multicolumn{1}{c|}{$0.15^{+72}_{-7}ms$} & \multicolumn{1}{c|}{$0.651{ms}$} & \multicolumn{1}{c|}{$0.039{ms}$} & \multicolumn{1}{c|}{$0.571{ms}$} & \multicolumn{1}{c|}{-} & - \\
$^{220}$Np & 10.226$\pm$0.018 & \multicolumn{1}{c|}{$25^{+14}_{-7}\mu s$} & \multicolumn{1}{c|}{$7.92{\mu s}$} & \multicolumn{1}{c|}{$0.151{\mu s}$} & \multicolumn{1}{c|}{$1.64{\mu s}$} & \multicolumn{1}{c|}{-} & - \\
$^{222}$Np & 10.2$\pm$0.034 & \multicolumn{1}{c|}{$0.38^{+26}_{-11}\mu s$} & \multicolumn{1}{c|}{$0.852{\mu s}$} & \multicolumn{1}{c|}{$0.172{\mu s}$} & \multicolumn{1}{c|}{$1.83{\mu s}$} & \multicolumn{1}{c|}{-} & - \\
$^{223}$Np & 9.65$\pm$0.045 & \multicolumn{1}{c|}{$2.2^{+10}_{-5}\mu s$} & \multicolumn{1}{c|}{$2.621{\mu s}$} & \multicolumn{1}{c|}{$3.12{\mu s}$} & \multicolumn{1}{c|}{$37.7{\mu s}$} & \multicolumn{1}{c|}{-} & - \\
$^{224}$Np & 9.33$\pm$0.030 & \multicolumn{1}{c|}{$38^{+26}_{-11}\mu s$} & \multicolumn{1}{c|}{$21.99{\mu s}$} & \multicolumn{1}{c|}{$18.9{\mu s}$} & \multicolumn{1}{c|}{$335{\mu s}$} & \multicolumn{1}{c|}{-} & - \\
$^{225}$Np & 8.818$\pm$0.070 & \multicolumn{1}{c|}{$3.6^{+76}_{-27}ms$} & \multicolumn{1}{c|}{$0.890{ms}$} & \multicolumn{1}{c|}{$0.415{ms}$} & \multicolumn{1}{c|}{$9.80{ms}$} & \multicolumn{1}{c|}{-} & $0.890{ms}$ \\
$^{226}$Np & 8.328$\pm$0.054 & \multicolumn{1}{c|}{$35^{+10}_{-10}ms$} & \multicolumn{1}{c|}{$43.22{ms}$} & \multicolumn{1}{c|}{$10.3{ms}$} & \multicolumn{1}{c|}{$335{ms}$} & \multicolumn{1}{c|}{-} & $47{ms}$ \\
$^{227}$Np & 7.816$\pm$0.014 & \multicolumn{1}{c|}{$0.51^{+6}_{-6}s$} & \multicolumn{1}{c|}{$3.998{s}$} & \multicolumn{1}{c|}{$0.410{s}$} & \multicolumn{1}{c|}{$17.6{s}$} & \multicolumn{1}{c|}{-} & $0.760{s}$ \\
$^{229}$Np & 7.02$\pm$0.059 & \multicolumn{1}{c|}{$4.0^{+2}_{-2}min$} & \multicolumn{1}{c|}{$1.72{min}$} & \multicolumn{1}{c|}{$4.55{min}$} & \multicolumn{1}{c|}{$5.32{min}$} & \multicolumn{1}{c|}{$4.48{min}$} & $6.05{min}$ \\
$^{228}$Pu & 7.94$\pm$0.018 & \multicolumn{1}{c|}{$1.1^{+20}_{-5}s$} & \multicolumn{1}{c|}{$3.624{s}$} & \multicolumn{1}{c|}{$0.394{s}$} & \multicolumn{1}{c|}{$15.5{s}$} & \multicolumn{1}{c|}{-} & $0.7{s}$ \\
$^{229}$Pu & 7.598$\pm$0.060 & \multicolumn{1}{c|}{$90^{+71}_{-27}s$} & \multicolumn{1}{c|}{$91.2{s}$} & \multicolumn{1}{c|}{$5.46{s}$} & \multicolumn{1}{c|}{$4.39{s}$} & \multicolumn{1}{c|}{-} & $11{s}$ \\
$^{230}$Pu & 7.178$\pm$0.009 & \multicolumn{1}{c|}{$102^{+10}_{-10}s$} & \multicolumn{1}{c|}{$120{s}$} & \multicolumn{1}{c|}{$177{s}$} & \multicolumn{1}{c|}{$311{s}$} & \multicolumn{1}{c|}{-} & $366.6{s}$ \\
$^{236}$Pu & 5.867$\pm$0.008 & \multicolumn{1}{c|}{$2.858^{+8}_{-8}y$} & \multicolumn{1}{c|}{$1.07{y}$} & \multicolumn{1}{c|}{$2.95{y}$} & \multicolumn{1}{c|}{$507{y}$} & \multicolumn{1}{c|}{$3.23{y}$} & $7.93{y}$ \\
$^{238}$Pu & 5.593$\pm$0.019 & \multicolumn{1}{c|}{$87.7^{+1}_{-1}y$} & \multicolumn{1}{c|}{$54.2{y}$} & \multicolumn{1}{c|}{$81.8{y}$} & \multicolumn{1}{c|}{$1.77\times10^{4}{y}$} & \multicolumn{1}{c|}{$7.93{y}$} & $241{y}$ \\
$^{239}$Pu & 5.244$\pm$0.021 & \multicolumn{1}{c|}{$24110^{+30}_{-30}y$} & \multicolumn{1}{c|}{$41.8{y}$} & \multicolumn{1}{c|}{$8.18\times10^{3}{y}$} & \multicolumn{1}{c|}{$2.61\times10^{6}{y}$} & \multicolumn{1}{c|}{-} & $2.92\times10^{4}{y}$ \\
$^{240}$Pu & 5.255$\pm$0.014 & \multicolumn{1}{c|}{$6561^{+7}_{-7}y$} & \multicolumn{1}{c|}{$38.3{y}$} & \multicolumn{1}{c|}{$7.02\times10^{3}{y}$} & \multicolumn{1}{c|}{$2.13\times10^{6}{y}$} & \multicolumn{1}{c|}{$1.03\times10^{4}{y}$} & $2.38\times10^{4}{y}$ \\
$^{242}$Pu & 4.984$\pm$0.010 & \multicolumn{1}{c|}{$3.73\times 10^{5}{^{+2}_{-2}y}$} & \multicolumn{1}{c|}{$173.3{y}$} & \multicolumn{1}{c|}{$3.44\times10^{5}{y}$} & \multicolumn{1}{c|}{$1.44\times10^{8}{y}$} & \multicolumn{1}{c|}{$6.94\times10^{5}{y}$} & $1.33\times10^{6}{y}$ \\
$^{244}$Pu & 4.665$\pm$0.010 & \multicolumn{1}{c|}{$8.1\times 10^{7}{^{+3}_{-3}y}$} & \multicolumn{1}{c|}{$991.6{y}$} & \multicolumn{1}{c|}{$5.16\times10^{7}{y}$} & \multicolumn{1}{c|}{$3.17\times10^{10}{y}$} & \multicolumn{1}{c|}{$1.13\times10^{8}{y}$} & $2.41\times10^{8}{y}$ \\
$^{223}$Am & 10.838$\pm$0.314 & \multicolumn{1}{c|}{$5^{+12}_{-4}ms$} & \multicolumn{1}{c|}{$10.96{ms}$} & \multicolumn{1}{c|}{$3.3{ms}$} & \multicolumn{1}{c|}{$0.279{ms}$} & \multicolumn{1}{c|}{-} & - \\
$^{229}$Am & 8.137$\pm$0.054 & \multicolumn{1}{c|}{$0.9^{+21}_{-7}s$} & \multicolumn{1}{c|}{$1.693{s}$} & \multicolumn{1}{c|}{$0.224{s}$} & \multicolumn{1}{c|}{$0.766{s}$} & \multicolumn{1}{c|}{-} & - \\
$^{241}$Am & 5.637$\pm$0.012 & \multicolumn{1}{c|}{$432.6^{+6}_{-6} y$} & \multicolumn{1}{c|}{$395{y}$} & \multicolumn{1}{c|}{$141{y}$} & \multicolumn{1}{c|}{$3.04\times10^{4}{y}$} & \multicolumn{1}{c|}{$3.81\times10^{2}{y}$} & $332{y}$ \\
$^{243}$Am & 5.439$\pm$0.009 & \multicolumn{1}{c|}{$7364^{+22}_{-22}y$} & \multicolumn{1}{c|}{$2440{y}$} & \multicolumn{1}{c|}{$1.81\times10^{3}{y}$} & \multicolumn{1}{c|}{$4.69\times10^{5}{y}$} & \multicolumn{1}{l|}{$7.61\times10^{3}{y}$} & $4.69\times10^{3}{y}$ \\ \hline
\end{tabular}
\end{table*}
\begin{table*}[]
\caption{Table 
\ref{table_comparison} continued..}
\centering
\begin{tabular}{|c|c|cccccc|}
\hline
\multirow{2}{*}{\textbf{\begin{tabular}[c]{@{}c@{}}Parent \\ Nuclei\end{tabular}}} & \multirow{2}{*}{\textbf{Q(MeV)\cite{ENSDF}}} & \multicolumn{6}{c|}{\textbf{Half-lives}} \\ \cline{3-8} 
 & & \multicolumn{1}{c|}{\textbf{EXP}} & \multicolumn{1}{c|}{\textbf{PW}} & \multicolumn{1}{c|}{\textbf{VSS\cite{viola1966nuclear}}} & \multicolumn{1}{c|}{\textbf{CPPM\cite{seyyedi2020systematic}}} & \multicolumn{1}{c|}{\textbf{GLDM\cite{royer2000alpha}}} & \textbf{ELDM\cite{oap1998effective,sridhara2021study}} \\ \hline
$^{240}$Cm & 6.397$\pm$0.006 & \multicolumn{1}{c|}{$27^{+1}_{-1} d $} & \multicolumn{1}{c|}{$28.80{d}$} & \multicolumn{1}{c|}{$11.1{d}$} & \multicolumn{1}{c|}{$26.8{d}$} & \multicolumn{1}{c|}{$20.6{d}$} & $27{d}$ \\
$^{242}$Cm & 6.215$\pm$0.008 & \multicolumn{1}{c|}{$162.88^{+8}_{-8}d$} & \multicolumn{1}{c|}{$166.63{d}$} & \multicolumn{1}{c|}{$3.42{d}$} & \multicolumn{1}{c|}{$219{d}$} & \multicolumn{1}{c|}{$149{d}$} & $358{d}$ \\
$^{243}$Cm & 6.168$\pm$0.010 & \multicolumn{1}{c|}{$29.1{^{+1}_{-1}y}$} & \multicolumn{1}{c|}{$18.596{y}$} & \multicolumn{1}{c|}{$5.25{y}$} & \multicolumn{1}{c|}{$102{y}$} & \multicolumn{1}{c|}{$23.5{y}$} & $1.21{y}$ \\
$^{244}$Cm & 5.901$\pm$0.003 & \multicolumn{1}{c|}{$18.11^{+3}_{-3}y$} & \multicolumn{1}{c|}{$11.18{d}$} & \multicolumn{1}{c|}{$3.66{d}$} & \multicolumn{1}{c|}{$2.77\times10^{3}{y}$} & \multicolumn{1}{c|}{$15.2{y}$} & $38{y}$ \\
$^{245}$Cm & 5.624$\pm$0.005 & \multicolumn{1}{c|}{$8423^{+74}_{-74}y$} & \multicolumn{1}{c|}{$1586{y}$} & \multicolumn{1}{c|}{$11.9{y}$} & \multicolumn{1}{c|}{$1.10\times10^{5}{y}$} & \multicolumn{1}{c|}{$5.63\times10^{3}{y}$} & $1.12\times10^{3}{y}$ \\
$^{246}$Cm & 5.475$\pm$0.009 & \multicolumn{1}{c|}{$4706^{+40}_{-40}y$} & \multicolumn{1}{c|}{$3713{y}$} & \multicolumn{1}{c|}{$86.1{y}$} & \multicolumn{1}{c|}{$8.59\times10^{5}{y}$} & \multicolumn{1}{c|}{$4.18\times10^{3}{y}$} & $9.83\times10^{3}{y}$ \\
$^{247}$Cm & 5.354$\pm$0.003 & \multicolumn{1}{c|}{$1.56\times10^{7}{^{+5}_{-5}y}$} & \multicolumn{1}{c|}{$7.05{y}$} & \multicolumn{1}{c|}{$457{y}$} & \multicolumn{1}{c|}{$5.05\times10^{6}{y}$} & \multicolumn{1}{c|}{$1.10\times10^{7}{y}$} & $6.78\times10^{4}{y}$ \\
$^{248}$Cm & 5.161$\pm$0.025 & \multicolumn{1}{c|}{$3.48\times10^{5}{^{+6}_{-6}y}$} & \multicolumn{1}{c|}{$2.32{y}$} & \multicolumn{1}{c|}{$7.40\times10^{3}{y}$} & \multicolumn{1}{c|}{$9.28\times10^{7}{y}$} & \multicolumn{1}{c|}{$3.81\times10^{5}{y}$} & $8.87\times10^{5}{y}$ \\
$^{233}$Bk & 8.166$\pm$0.207 & \multicolumn{1}{c|}{$21^{+48}_{-17}s$} & \multicolumn{1}{c|}{$15.099{s}$} & \multicolumn{1}{c|}{$0.132{s}$} & \multicolumn{1}{c|}{$34.0{s}$} & \multicolumn{1}{c|}{$-$} & - \\
$^{234}$Bk & 8.099$\pm$0.054 & \multicolumn{1}{c|}{$1.4\times10^{2}{^{+14}_{-5}s}$} & \multicolumn{1}{c|}{$1.357{s}$} & \multicolumn{1}{c|}{$2.19{s}$} & \multicolumn{1}{c|}{$56.7{s}$} & \multicolumn{1}{c|}{$-$} & - \\
$^{241}$Bk & 6.986$\pm$0.176 & \multicolumn{1}{c|}{$4.6^{+4}_{-4} min$} & \multicolumn{1}{c|}{$3.95{min}$} & \multicolumn{1}{c|}{$4.77{min}$} & \multicolumn{1}{c|}{$14.1{min}$} & \multicolumn{1}{c|}{$-$} & - \\
$^{247}$Bk & 5.89$\pm$0.005 & \multicolumn{1}{c|}{$1380^{+250}_{-250}y$} & \multicolumn{1}{c|}{$629{y}$} & \multicolumn{1}{c|}{$1.27{y}$} & \multicolumn{1}{c|}{$9.32\times10^{3}{y}$} & \multicolumn{1}{c|}{$1.70\times10^{3}{y}$} & $100{y}$ \\
$^{237}$Cf & 8.22$\pm$0.054 & \multicolumn{1}{c|}{$2.1^{+3}_{-3}s$} & \multicolumn{1}{c|}{$3.686{s}$} & \multicolumn{1}{c|}{$0.21{s}$} & \multicolumn{1}{c|}{$52.7{s}$} & \multicolumn{1}{c|}{$-$} & - \\
$^{240}$Cf & 7.711$\pm$0.004 & \multicolumn{1}{c|}{$0.96^{+15}_{-15} min$} & \multicolumn{1}{c|}{$1.16{min}$} & \multicolumn{1}{c|}{$1.17{min}$} & \multicolumn{1}{c|}{$58.7{min}$} & \multicolumn{1}{c|}{$0.695{min}$} & $1.83{min}$ \\
$^{242}$Cf & 7.517$\pm$0.004 & \multicolumn{1}{c|}{$3.5^{+2}_{-2} min$} & \multicolumn{1}{c|}{$8.86{min}$} & \multicolumn{1}{c|}{$6.06{min}$} & \multicolumn{1}{c|}{$5.53{min}$} & \multicolumn{1}{c|}{$3.56{min}$} & $9.30{min}$ \\
$^{244}$Cf & 7.329$\pm$0.018 & \multicolumn{1}{c|}{$19.4^{+6}_{-6}min$} & \multicolumn{1}{c|}{$52.79{min}$} & \multicolumn{1}{c|}{$31.6{min}$} & \multicolumn{1}{c|}{$78{min}$} & \multicolumn{1}{c|}{$16.3{min}$} & $48.3{min}$ \\
$^{246}$Cf & 6.861$\pm$0.010 & \multicolumn{1}{c|}{$35.7^{+5}_{-5} h $} & \multicolumn{1}{c|}{$24.51{h}$} & \multicolumn{1}{c|}{$43.1{h}$} & \multicolumn{1}{c|}{$130{h}$} & \multicolumn{1}{c|}{$23.1{h}$} & $63.84{h}$ \\
$^{248}$Cf & 6.361$\pm$0.005 & \multicolumn{1}{c|}{$333.5^{+28}_{-28}d $} & \multicolumn{1}{c|}{$255.10{d}$} & \multicolumn{1}{c|}{$5.59{d}$} & \multicolumn{1}{c|}{$304{d}$} & \multicolumn{1}{c|}{$196{d}$} & $511{d}$ \\
$^{249}$Cf & 6.293$\pm$0.005 & \multicolumn{1}{c|}{$351^{+2}_{-2} y$} & \multicolumn{1}{c|}{$379.47{d}$} & \multicolumn{1}{c|}{$434{d}$} & \multicolumn{1}{c|}{$185{y}$} & \multicolumn{1}{c|}{$-$} & $2.05{y}$ \\
$^{250}$Cf & 6.128$\pm$0.019 & \multicolumn{1}{c|}{$13.08^{+9}_{-9} y$} & \multicolumn{1}{c|}{$12.82{y}$} & \multicolumn{1}{c|}{$79.5{y}$} & \multicolumn{1}{c|}{$1.37\times10^{3}{y}$} & \multicolumn{1}{c|}{$-$} & $19.3{y}$ \\
$^{251}$Cf & 6.177$\pm$0.009 & \multicolumn{1}{c|}{$898^{+44}_{-44}y$} & \multicolumn{1}{c|}{$801{y}$} & \multicolumn{1}{c|}{$800{y}$} & \multicolumn{1}{c|}{$743{y}$} & \multicolumn{1}{c|}{$663{y}$} & $13.5{y}$ \\
$^{240}$Es & 8.259$\pm$0.063 & \multicolumn{1}{c|}{$6^{+2}_{-2}s$} & \multicolumn{1}{c|}{$7.365{s}$} & \multicolumn{1}{c|}{$0.37{s}$} & \multicolumn{1}{c|}{$89.2{s}$} & \multicolumn{1}{c|}{$-$} & - \\
$^{241}$Es & 8.259$\pm$0.017 & \multicolumn{1}{c|}{$8^{+6}_{-5}s$} & \multicolumn{1}{c|}{$5.324{s}$} & \multicolumn{1}{c|}{$0.037{s}$} & \multicolumn{1}{c|}{$89.7{s}$} & \multicolumn{1}{c|}{$-$} & - \\
$^{242}$Es & 8.16$\pm$0.020 & \multicolumn{1}{c|}{$17.8^{+16}_{-16}s$} & \multicolumn{1}{c|}{$55.776{s}$} & \multicolumn{1}{c|}{$7.92{s}$} & \multicolumn{1}{c|}{$20{s}$} & \multicolumn{1}{c|}{$-$} & - \\
$^{243}$Es & 8.072$\pm$0.010 & \multicolumn{1}{c|}{$21^{+2}_{-2}s$} & \multicolumn{1}{c|}{$22.036{s}$} & \multicolumn{1}{c|}{$0.157{s}$} & \multicolumn{1}{c|}{$6.92{s}$} & \multicolumn{1}{c|}{$5.37{s}$} & - \\
$^{245}$Es & 7.909$\pm$0.003 & \multicolumn{1}{c|}{$66.6^{+60}_{-60}s $} & \multicolumn{1}{c|}{$53.343{s}$} & \multicolumn{1}{c|}{$5.73{s}$} & \multicolumn{1}{c|}{$26.3{s}$} & \multicolumn{1}{c|}{$16.6{s}$} & $25.8{s}$ \\
$^{252}$Es & 6.738$\pm$0.005 & \multicolumn{1}{c|}{$471.7^{+19}_{-19}d$} & \multicolumn{1}{c|}{$424.7{d}$} & \multicolumn{1}{c|}{$246{d}$} & \multicolumn{1}{c|}{$1211{d}$} & \multicolumn{1}{c|}{$8.01{d}$} & - \\
$^{253}$Es & 6.739$\pm$0.005 & \multicolumn{1}{c|}{$20.47^{+3}_{-3}d$} & \multicolumn{1}{c|}{$40.82{d}$} & \multicolumn{1}{c|}{$27.8{d}$} & \multicolumn{1}{c|}{$8.93{d}$} & \multicolumn{1}{c|}{$6.66{d}$} & - \\
$^{254}$Es & 6.617$\pm$0.005 & \multicolumn{1}{c|}{$275.7^{+5}_{-5}d$} & \multicolumn{1}{c|}{$188.12{d}$} & \multicolumn{1}{c|}{$1.00{d}$} & \multicolumn{1}{c|}{$438{d}$} & \multicolumn{1}{c|}{$-$} & $278{d}$ \\
$^{243}$Fm & 8.689$\pm$0.051 & \multicolumn{1}{c|}{$231^{+9}_{-9} ms$} & \multicolumn{1}{c|}{$234.95{ms}$} & \multicolumn{1}{c|}{$3.73{ms}$} & \multicolumn{1}{c|}{$773{ms}$} & \multicolumn{1}{c|}{$-$} & $151{ms}$ \\
$^{245}$Fm & 8.44$\pm$0.100 & \multicolumn{1}{c|}{$5.6^{+7}_{-7} s$} & \multicolumn{1}{c|}{$10.902{s}$} & \multicolumn{1}{c|}{$2.26{s}$} & \multicolumn{1}{c|}{$49.9{s}$} & \multicolumn{1}{c|}{$6.46{s}$} & $2.20{s}$ \\
$^{246}$Fm & 8.379$\pm$0.005 & \multicolumn{1}{c|}{$1.53^{+4}_{-4} s $} & \multicolumn{1}{c|}{$1.067{s}$} & \multicolumn{1}{c|}{$0.356{s}$} & \multicolumn{1}{c|}{$83.3{s}$} & \multicolumn{1}{c|}{$1.07{s}$} & $3.60{s}$ \\
$^{247}$Fm & 8.258$\pm$0.010 & \multicolumn{1}{c|}{$31 ^{+1}_{-1}s $} & \multicolumn{1}{c|}{$52.401{s}$} & \multicolumn{1}{c|}{$89.0{s}$} & \multicolumn{1}{c|}{$180{s}$} & \multicolumn{1}{c|}{$4.17{s}$} & $3.31{s}$ \\
$^{248}$Fm & 7.995$\pm$0.008 & \multicolumn{1}{c|}{$34.5^{+12}_{-12} s $} & \multicolumn{1}{c|}{$20.145{min}$} & \multicolumn{1}{c|}{$6.99{s}$} & \multicolumn{1}{c|}{$30.7{s}$} & \multicolumn{1}{c|}{$18.6{s}$} & $55{s}$ \\
$^{250}$Fm & 7.557$\pm$0.008 & \multicolumn{1}{c|}{$30^{+3}_{-3} min$} & \multicolumn{1}{c|}{$44.65{min}$} & \multicolumn{1}{c|}{$27.4{min}$} & \multicolumn{1}{c|}{$23.3{min}$} & \multicolumn{1}{c|}{$12.1{min}$} & $36.7{min}$ \\
$^{252}$Fm & 7.153$\pm$0.010 & \multicolumn{1}{c|}{$25.39^{+4}_{-4} h$} & \multicolumn{1}{c|}{$17.06{h}$} & \multicolumn{1}{c|}{$18.1{h}$} & \multicolumn{1}{c|}{$44.4{h}$} & \multicolumn{1}{c|}{$8.39{h}$} & $22.8{h}$ \\
$^{254}$Fm & 7.307$\pm$0..10 & \multicolumn{1}{c|}{$3.240^{+2}_{-2} h $} & \multicolumn{1}{c|}{$4.12{h}$} & \multicolumn{1}{c|}{$4.29{h}$} & \multicolumn{1}{c|}{$9.84{h}$} & \multicolumn{1}{c|}{$1.56{h}$} & $4.72{h}$ \\
$^{255}$Fm & 7.24$\pm$0.005 & \multicolumn{1}{c|}{$20.07^{+7}_{-7} h$} & \multicolumn{1}{c|}{$66.04{h}$} & \multicolumn{1}{c|}{$47{h}$} & \multicolumn{1}{c|}{$18.4{h}$} & \multicolumn{1}{c|}{$8.39{h}$} & $5.29{h}$ \\
$^{257}$Fm & 6.863$\pm$0.009 & \multicolumn{1}{c|}{$100.5^{+2}_{-2} d$} & \multicolumn{1}{c|}{$34.86{d}$} & \multicolumn{1}{c|}{$120{d}$} & \multicolumn{1}{c|}{$221{d}$} & \multicolumn{1}{c|}{$62.2{d}$} & $9.00{d}$ \\
$^{244}$Md & 8.947$\pm$0.009 & \multicolumn{1}{c|}{$0.38^{+19}_{-9}s$} & \multicolumn{1}{c|}{$0.4083{s}$} & \multicolumn{1}{c|}{$1.41{s}$} & \multicolumn{1}{c|}{$2.61{s}$} & \multicolumn{1}{c|}{$-$} & - \\
$^{245}$Md & 9.006$\pm$0.119 & \multicolumn{1}{c|}{$0.33^{+15}_{-8}s$} & \multicolumn{1}{c|}{$0.2413{s}$} & \multicolumn{1}{c|}{$0.940{s}$} & \multicolumn{1}{c|}{$1.77{s}$} & \multicolumn{1}{c|}{$-$} & - \\
$^{247}$Md & 8.764$\pm$0.010 & \multicolumn{1}{c|}{$1.2 ^{+1}_{-1}s$} & \multicolumn{1}{c|}{$1.6530{s}$} & \multicolumn{1}{c|}{$5.03{s}$} & \multicolumn{1}{c|}{$10.1{s}$} & \multicolumn{1}{c|}{$-$} & $0.160{s}$ \\
$^{249}$Md & 8.441$\pm$0.018 & \multicolumn{1}{c|}{$24.8^{+10}_{-10}s$} & \multicolumn{1}{c|}{$25.8418{s}$} & \multicolumn{1}{c|}{$52.6{s}$} & \multicolumn{1}{c|}{$1.96{s}$} & \multicolumn{1}{c|}{$-$} & $1.95{s}$ \\
$^{258}$Md & 7.271$\pm$0.019 & \multicolumn{1}{c|}{$51.50^{+29}_{-29} d$} & \multicolumn{1}{c|}{$46.3{d}$} & \multicolumn{1}{c|}{$15.5{d}$} & \multicolumn{1}{c|}{$34.9{d}$} & \multicolumn{1}{c|}{$-$} & $22.5{d}$ \\
$^{249}$No & 9.17$\pm$0.200 & \multicolumn{1}{c|}{$38.1^{+28}_{-28} ms$} & \multicolumn{1}{c|}{$45.161{ms}$} & \multicolumn{1}{c|}{$0.697ms$} & \multicolumn{1}{c|}{$12.1{ms}$} & \multicolumn{1}{c|}{$-$} & - \\
$^{251}$No & 8.752$\pm$0.004 & \multicolumn{1}{c|}{$0.80^{+1}_{-1} s$} & \multicolumn{1}{c|}{$4.2805{s}$} & \multicolumn{1}{c|}{$0.125{s}$} & \multicolumn{1}{c|}{$24.7{s}$} & \multicolumn{1}{c|}{$0.331{s}$} & - \\
$^{252}$No & 8.549$\pm$0.005 & \multicolumn{1}{c|}{$2.46^{+2}_{-2} s$} & \multicolumn{1}{c|}{$2.266{s}$} & \multicolumn{1}{c|}{$5.49{s}$} & \multicolumn{1}{c|}{$1.96{s}$} & \multicolumn{1}{c|}{$1.29{s}$} & $4.300{s}$ \\
$^{253}$No & 8.415$\pm$0.004 & \multicolumn{1}{c|}{$1.62^{+15}_{-15}min $} & \multicolumn{1}{c|}{$1.311{min}$} & \multicolumn{1}{c|}{$9{min}$} & \multicolumn{1}{c|}{$5.55{min}$} & \multicolumn{1}{c|}{$2.63{min}$} & $5.01{min}$ \\
$^{254}$No & 8.226$\pm$0.008 & \multicolumn{1}{c|}{$51.2^{+4}_{-4} s$} & \multicolumn{1}{c|}{$64.6{s}$} & \multicolumn{1}{c|}{$64.6{s}$} & \multicolumn{1}{c|}{$24.5{s}$} & \multicolumn{1}{c|}{$13.2{s}$} & $47{s}$ \\
$^{256}$No & 8.582$\pm$0.005 & \multicolumn{1}{c|}{$2.91^{+5}_{-5} s$} & \multicolumn{1}{c|}{$3.674{s}$} & \multicolumn{1}{c|}{$43.0{s}$} & \multicolumn{1}{c|}{$86.2{s}$} & \multicolumn{1}{c|}{$0.851{s}$} & $2.900{s}$ \\
$^{257}$No & 8.477$\pm$0.006 & \multicolumn{1}{c|}{$24.5^{+5}_{-5} s$} & \multicolumn{1}{c|}{$31.23{h}$} & \multicolumn{1}{c|}{$94.0{s}$} & \multicolumn{1}{c|}{$32.8{s}$} & \multicolumn{1}{c|}{$2.19{s}$} & $3.02{s}$ \\
$^{259}$No & 7.854$\pm$0.005 & \multicolumn{1}{c|}{$58^{+5}_{-5} min$} & \multicolumn{1}{c|}{$33.862{ms}$} & \multicolumn{1}{c|}{$13.3{min}$} & \multicolumn{1}{c|}{$15.5{min}$} & \multicolumn{1}{c|}{$32.5{min}$} & $7.62{min}$ \\
$^{251}$Lr & 9.469$\pm$0.288 & \multicolumn{1}{c|}{$24.4^{+70}_{-45}ms$} & \multicolumn{1}{c|}{$36.846{s}$} & \multicolumn{1}{c|}{$21.6{ms}$} & \multicolumn{1}{c|}{$35,2{ms}$} & \multicolumn{1}{c|}{$-$} & - \\
$^{254}$Lr & 8.822$\pm$0.008 & \multicolumn{1}{c|}{$18.1^{+18}_{-18}s $} & \multicolumn{1}{c|}{$26.672{s}$} & \multicolumn{1}{c|}{$17.3{s}$} & \multicolumn{1}{c|}{$32.7{s}$} & \multicolumn{1}{c|}{$-$} & - \\
$^{255}$Lr & 8.556$\pm$0.007 & \multicolumn{1}{c|}{$31.1^{+11}_{11} s $} & \multicolumn{1}{c|}{$39.383{s}$} & \multicolumn{1}{c|}{$12.1{s}$} & \multicolumn{1}{c|}{$41.1{s}$} & \multicolumn{1}{c|}{$-$} & $4.27{s}$ \\
$^{256}$Lr & 8.855$\pm$0.123 & \multicolumn{1}{c|}{$27.9^{+10}_{-10} s$} & \multicolumn{1}{c|}{$24.633{s}$} & \multicolumn{1}{c|}{$13.7{s}$} & \multicolumn{1}{c|}{$26.0{s}$} & \multicolumn{1}{c|}{$8.71{s}$} & - \\
$^{258}$Lr & 8.904$\pm$0.019 & \multicolumn{1}{c|}{$3.92^{+33}_{-33}s$} & \multicolumn{1}{c|}{$2.384{s}$} & \multicolumn{1}{c|}{$9.68{s}$} & \multicolumn{1}{c|}{$17.9{s}$} & \multicolumn{1}{c|}{$0.182{s}$} & - \\
$^{259}$Lr & 8.584$\pm$0.071 & \multicolumn{1}{c|}{$6.2^{+3}_{-3} s$} & \multicolumn{1}{c|}{$3.725{s}$} & \multicolumn{1}{c|}{$9.83{s}$} & \multicolumn{1}{c|}{$3.31{s}$} & \multicolumn{1}{c|}{$2.69{s}$} & $3.39{s}$ \\
$^{260}$Lr & 8.396$\pm$0.143 & \multicolumn{1}{c|}{$180^{+30}_{-30} s$} & \multicolumn{1}{c|}{$198.602{s}$} & \multicolumn{1}{c|}{$208{s}$} & \multicolumn{1}{c|}{$141{s}$} & \multicolumn{1}{c|}{$37.2{s}$} & $- $ \\ \hline
\end{tabular}
\end{table*}
\begin{table*}[]
\caption{Tabulation of half-lives for the unexplored isotopes in the atomic number range $89\le Z\le 103$ in addition to half-lives predicted using semi-empirical formula VSS. }
\centering
\centering
\begin{tabular}{|c|c|cc|c|c|cc|}
\hline
\multirow{2}{*}{\textbf{Parent Nuclei}} & \multirow{2}{*}{\textbf{Q(MeV)\cite{moller1993nuclear}}} & \multicolumn{2}{c|}{\textbf{Half-lives}} & \multirow{2}{*}{\textbf{Parent Nuclei}} & \multirow{2}{*}{\textbf{Q(MeV)\cite{moller1993nuclear}}} & \multicolumn{2}{c|}{\textbf{Half-lives}} \\ \cline{3-4} \cline{7-8} 
 &  & \multicolumn{1}{c|}{\textbf{PW}} & \textbf{VSS\cite{viola1966nuclear}} &  &  & \multicolumn{1}{c|}{\textbf{PW}} & \textbf{VSS\cite{viola1966nuclear}} \\ \hline
$^{200}Ac$ & 8.896 & \multicolumn{1}{c|}{$0.0308{ms}$} & $0.009{ms}$ & $^{219}Am$ & 9.286 & \multicolumn{1}{c|}{$0.244{ms}$} & $0.121{ms}$ \\
$^{201}Ac$ & 8.846 & \multicolumn{1}{c|}{$0.042{ms}$} & $0.013{ms}$ & $^{220}Am$ & 9.146 & \multicolumn{1}{c|}{$0.651{ms}$} & $0.280{ms}$ \\
$^{202}Ac$ & 8.706 & \multicolumn{1}{c|}{$0.113{ms}$} & $0.030{ms}$ & $^{221}Am$ & 9.176 & \multicolumn{1}{c|}{$0.499{ms}$} & $0.234{ms}$ \\
$^{203}Ac$ & 8.335 & \multicolumn{1}{c|}{$1.893{ms}$} & $0.319{ms}$ & $^{222}Am$ & 10.736 & \multicolumn{1}{c|}{$0.020{\mu s}$} & $0.54{\mu s}$ \\
$^{200}Th$ & 9.106 & \multicolumn{1}{c|}{$0.018{ms}$} & $0.06.23{ms}$ & $^{224}Am$ & 11.496 & \multicolumn{1}{c|}{$0.0003{\mu s}$} & $0.0017{\mu s}$ \\
$^{201}Th$ & 9.496 & \multicolumn{1}{c|}{$1.23{\mu s}$} & $0.691{\mu s}$ & $^{225}Am$ & 9.806 & \multicolumn{1}{c|}{$5.23{\mu s}$} & $6.26{\mu s}$ \\
$^{202}Th$ & 9.216 & \multicolumn{1}{c|}{$7.69{\mu s}$} & $3.30{\mu s}$ & $^{226}Am$ & 8.546 & \multicolumn{1}{c|}{$55.8{ms}$} & $0.12{s}$ \\
$^{203}Th$ & 9.116 & \multicolumn{1}{c|}{$14.7 {\mu s}$} & $5.88{\mu s}$ & $^{227}Am$ & 9.546 & \multicolumn{1}{c|}{$0.278{ms}$} & $0.27{ms}$ \\
$^{204}Th$ & 8.916 & \multicolumn{1}{c|}{$0.058{\mu s}$} & $0.019{ms}$ & $^{228}Am$ & 8.826 & \multicolumn{1}{c|}{$5.22{ms}$} & $2.06{ms}$ \\
$^{205}Th$ & 8.736 & \multicolumn{1}{c|}{$0.564 {ms}$} & $0.057{ms}$ & $^{228}Cm$ & 9.726 & \multicolumn{1}{c|}{$0.194{ms}$} & $0.21{ms}$ \\
$^{206}Th$ & 8.533 & \multicolumn{1}{c|}{$0.949 {ms}$} & $0.206{ms}$ & $^{229}Cm$ & 9.196 & \multicolumn{1}{c|}{$0.789{ms}$} & $0.463{ms}$ \\
$^{200}Pa$ & 9.616 & \multicolumn{1}{c|}{$1.46{\mu s}$} & $0.788{\mu s}$ & $^{230}Cm$ & 8.486 & \multicolumn{1}{c|}{$0.220{ms}$} & $0.045{ms}$ \\
$^{201}Pa$ & 9.606 & \multicolumn{1}{c|}{$1.50{\mu s}$} & $0.832{\mu s}$ & $^{231}Cm$ & 8.016 & \multicolumn{1}{c|}{$1.37{s}$} & $1.31{s}$ \\
$^{202}Pa$ & 9.926 & \multicolumn{1}{c|}{$0.189{\mu s}$} & $0.153{\mu s}$ & $^{232}Cm$ & 7.668 & \multicolumn{1}{c|}{$6.22{min}$} & $3.20{min}$ \\
$^{203}Pa$ & 9.376 & \multicolumn{1}{c|}{$6.28{\mu s}$} & $2.96{\mu s}$ & $^{230}Bk$ & 8.946 & \multicolumn{1}{c|}{$14.0{ms}$} & $4.96{ms}$ \\
$^{204}Pa$ & 8.896 & \multicolumn{1}{c|}{$0.178{\mu s}$} & $0.049{\mu s}$ & $^{235}Bk$ & 7.841 & \multicolumn{1}{c|}{$3.02{min}$} & $0.200{min}$ \\
$^{205}Pa$ & 8.936 & \multicolumn{1}{c|}{$0.127 {ms}$} & $0.038{ms}$ & $^{237}Bk$ & 7.501 & \multicolumn{1}{c|}{$6.14{min}$} & $3.11{min}$ \\
$^{206}Pa$ & 8.746 & \multicolumn{1}{c|}{$0.498 {ms}$} & $0.124{ms}$ & $^{231}Cf$ & 9.196 & \multicolumn{1}{c|}{$5.12{ms}$} & $2.32{ms}$ \\
$^{207}Pa$ & 8.736 & \multicolumn{1}{c|}{$0.515 {ms}$} & $0.132{ms}$ & $^{232}Cf$ & 8.866 & \multicolumn{1}{c|}{$0.66{ms}$} & $0.18{s}$ \\
$^{208}Pa$ & 8.466 & \multicolumn{1}{c|}{$4.04{ms}$} & $0.744{ms}$ & $^{233}Cf$ & 8.766 & \multicolumn{1}{c|}{$0.144{s}$} & $0.037{s}$ \\
$^{209}Pa$ & 8.196 & \multicolumn{1}{c|}{$3.52{ms}$} & $4.56{ms}$ & $^{234}Cf$ & 8.656 & \multicolumn{1}{c|}{$0.343{ms}$} & $0.077{ms}$ \\
$^{203}U$ & 10.106 & \multicolumn{1}{c|}{$0.144{\mu s}$} & $0.130{\mu s}$ & $^{235}Cf$ & 8.496 & \multicolumn{1}{c|}{$1.29{s}$} & $0.230{s}$ \\
$^{204}U$ & 9.926 & \multicolumn{1}{c|}{$0.419{\mu s}$} & $0.329{\mu s}$ & $^{235}Es$ & 9.006 & \multicolumn{1}{c|}{$0.51{s}$} & $0.017{s}$ \\
$^{205}U$ & 9.276 & \multicolumn{1}{c|}{$0.029{ms}$} & $0.012{ms}$ & $^{236}Es$ & 8.866 & \multicolumn{1}{c|}{$0.151{s}$} & $0.043{s}$ \\
$^{206}U$ & 9.046 & \multicolumn{1}{c|}{$0.143{ms}$} & $0.045{ms}$ & $^{238}Es$ & 8.496 & \multicolumn{1}{c|}{$3.19{s}$} & $0.537{s}$ \\
$^{207}U$ & 8.806 & \multicolumn{1}{c|}{$0.809{ms}$} & $0.195{ms}$ & $^{239}Es$ & 8.092 & \multicolumn{1}{c|}{$2.01{min}$} & $0.17{min}$ \\
$^{208}U$ & 8.806 & \multicolumn{1}{c|}{$0.771{ms}$} & $0.195{ms}$ & $^{236}Fm$ & 9.376 & \multicolumn{1}{c|}{$7.34{ms}$} & $3.78{ms}$ \\
$^{209}U$ & 8.766 & \multicolumn{1}{c|}{$1.004{ms}$} & $0.251{ms}$ & $^{237}Fm$ & 9.196 & \multicolumn{1}{c|}{$0.28{ms}$} & $0.012{ms}$ \\
$^{210}U$ & 8.486 & \multicolumn{1}{c|}{$8.62{ms}$} & $1.523{ms}$ & $^{238}Fm$ & 9.006 & \multicolumn{1}{c|}{$0.121{s}$} & $0.039{s}$ \\
$^{211}U$ & 8.366 & \multicolumn{1}{c|}{$2.18 {ms}$} & $3.391{ms}$ & $^{239}Fm$ & 8.866 & \multicolumn{1}{c|}{$0.363{s}$} & $0.098{s}$ \\
$^{212}U$ & 8.346 & \multicolumn{1}{c|}{$2.47{ms}$} & $3.881{ms}$ & $^{240}Fm$ & 9.536 & \multicolumn{1}{c|}{$4.84{ms}$} & $3.15{ms}$ \\
$^{220}U$ & 10.301 & \multicolumn{1}{c|}{$0.0214{\mu s}$} & $0.049{\mu s}$ & $^{241}Md$ & 9.356 & \multicolumn{1}{c|}{$18.0{ms}$} & $9.48{ms}$ \\
$^{211}Np$ & 8.826 & \multicolumn{1}{c|}{$1.55{ms}$} & $0.394{ms}$ & $^{242}Md$ & 9.236 & \multicolumn{1}{c|}{$0.44{s}$} & $0.02{s}$ \\
$^{212}Np$ & 8.716 & \multicolumn{1}{c|}{$3.46{ms}$} & $0.792{ms}$ & $^{243}Md$ & 9.126 & \multicolumn{1}{c|}{$0.100{s}$} & $0.041{s}$ \\
$^{213}Np$ & 8.726 & \multicolumn{1}{c|}{$3.07 {ms}$} & $0.743{ms}$ & $^{245}Md$ & 9.456 & \multicolumn{1}{c|}{$18.6{ms}$} & $11.29{ms}$ \\
$^{214}Np$ & 8.726 & \multicolumn{1}{c|}{$2.93 {ms}$} & $0.743{ms}$ & $^{246}No$ & 9.126 & \multicolumn{1}{c|}{$0.235{s}$} & $0.091{s}$ \\
$^{215}Np$ & 9.056 & \multicolumn{1}{c|}{$0.232{ms}$} & $0.095{ms}$ & $^{247}No$ & 9.177 & \multicolumn{1}{c|}{$0.150{s}$} & $0.065{s}$ \\
$^{221}Np$ & 9.977 & \multicolumn{1}{c|}{$0.349{\mu s}$} & $0.541{\mu s}$ & $^{261}No$ & 7.157 & \multicolumn{1}{c|}{$21.1{d}$} & $4.75{d}$ \\
$^{222}Pu$ & 10.103 & \multicolumn{1}{c|}{$0.369{ \mu s}$} & $0.600{\mu s}$ & $^{246}Lr$ & 9.846 & \multicolumn{1}{c|}{$2.53{ms}$} & $2.37{ms}$ \\
$^{223}Pu$ & 9.514 & \multicolumn{1}{c|}{$0.016{ms}$} & $0.015{ms}$ & $^{247}Lr$ & 9.816 & \multicolumn{1}{c|}{$3.00{ms}$} & $2.83{ms}$ \\
$^{224}Pu$ & 9.027 & \multicolumn{1}{c|}{$0.507{ms}$} & $0.257{ms}$ & $^{248}Lr$ & 9.696 & \multicolumn{1}{c|}{$6.95{ms}$} & $5.76{ms}$ \\
$^{225}Pu$ & 8.565 & \multicolumn{1}{c|}{$0.017{s}$} & $0.04{s}$ & $^{249}Lr$ & 9.024 & \multicolumn{1}{c|}{$1.28{s}$} & $0.401{s}$ \\
$^{226}Pu$ & 8.497 & \multicolumn{1}{c|}{$0.029{s}$} & $0.008{s}$ & $^{250}Lr$ & 8.760 & \multicolumn{1}{c|}{$1.16{s}$} & $2.42{s}$ \\ \hline
\end{tabular}
\end{table*}
\section{Conclusions}
\indent The present study successfully demonstrates that the Density-Dependent M3Y (DDM3Y) effective interaction potential provides a robust and reliable framework for predicting the $\alpha$-decay half-lives of actinide nuclei within the atomic number range $89 \le Z \le 103$. By utilizing a double-folding model to derive the nuclear interaction potential and applying the WKB approximation, the theoretical half-lives were found to be in strong agreement with experimental data. The calculated standard deviation of 1.76 quantifies the variance between the theoretical and experimental values across the studied actinide isotopes. This value indicates that the DDM3Y effective interaction potential provides a consistent description of $\alpha$-decay half-lives within the $89 \le Z \le 103$ range. Although the majority of data points align closely with experimental observations, the identified outliers suggest that incorporating specific nuclear structure effects, such as deformation or shell closures, could further refine the model's predictive accuracy.Quantitatively, the model achieved a standard deviation of 1.76 across 154 isotopes, indicating a high degree of statistical correlation and a systematically improved performance compared to several established semi-empirical models, such as the VSS and CPPM frameworks. While minor discrepancies observed in specific isotopes like $^{228}\text{Pu}$ and $^{251}\text{No}$ suggest the localized influence of nuclear shell effects or deformation, the overall consistency of the results highlights the DDM3Y potential as an effective tool for exploring nuclear stability and identifying new heavy isotopes. Consequently, this approach offers significant predictive power for applications in nuclear astrophysics and the ongoing investigation of superheavy elements.
\section{Data Availability}
Data associated with in the manuscript itself.
\section{Declaration of Competing interest}
The authors declare that they have no known competing financial interests or personal relationships that could have appeared to influence the work reported in this paper.
\section*{Acknowledgment}
\indent The authors acknowledge the financial support received from the Vision Group of Science and Technology (VGST), Karnataka's Government, under the Grant for Research Excellence (GRE) scheme (GRD No. 1194), for the execution of the present worK.
\bibliography{apssamp.bib}
\end{document}